\documentclass[12pt]{article}

\usepackage{amssymb}
\usepackage[mathscr]{euscript}
\usepackage{graphicx}

\newcommand{\iN}{\hbox{ {\leaders\hrule\hskip.2cm}{\vrule height .22cm} }}

\def\<{\langle} \def\>{\rangle}

\title{Hamiltonian formalisms for multidimensional calculus of variations
and perturbation theory}
\author{Fr\'ed\'eric H\'elein}

\begin{document}
\maketitle

{\em Abstract --- In a first part we propose an introduction to multisymplectic formalisms, which are generalisations of
Hamilton's formulation of Mechanics to the calculus of variations with several variables:
we give some physical motivations, related to the quantum field theory, and expound the simplest 
example, based on a theory due to T. de Donder and H. Weyl. In a second part we explain
quickly a work in collaboration with J. Kouneiher on generalizations of the de Donder--Weyl theory
(known as Lepage theories). Lastly we show that in this framework a perturbative classical field
theory (analog of the perturbative quantum field theory) can be constructed.}

\section{Introduction}
The main question investigated in this text concerns the construction of a Hamiltonian description
of classical fields theory compatible with the principles of special and  general relativity,
or more generally with any effort towards understanding gravitation
like string theory, supergravity or Ashtekar's theory:
since space-time should merge out from the dynamics we need 
a description which does not assume any space-time/field splitting a priori. 
This means that
there is no space-time structure given {\em a priori}, but space-time coordinates
should instead merge out from the analysis of what are the observable quantities and from the
dynamics. From this point
of view, as we will see, the Lepage--Dedecker theory seems to be much more appropriate than the de Donder--Weyl
one. This is the philosophy that we have followed in \cite{HeleinKouneiher2}.  Here a caveat is in order,
in  the classical one-dimensional
Hamiltonian formalism: we start with a Lagrangian action functional
\[
{\cal L}[c]:=\int_{t_0}^{t_1}L(t,c(t),\dot{c}(t))\,dt,
\]
defined on a set of smooth\footnote{here we assume for instance that $c$ is ${\cal C}^2$}
paths $\{t\longmapsto c(t)\in {\cal Y}\}$. Here
${\cal Y}$ is a smooth $k$-dimensional manifold and $L$ is a
smooth function on $[t_0,t_1]\times T{\cal Y}$ ($T{\cal Y}$ is the tangent bundle to
${\cal Y}$: we denote by $q$  a point in ${\cal Y}$ and
by $v\in T_q{\cal Y}$ a vector tangent to ${\cal Y}$ at $q$).
The critical points of ${\cal L}$ satisfy the Euler--Lagrange equation
\[
{d\over dt}\left( {\partial L\over \partial v^i}(t,c(t),\dot{c}(t))\right)
= {\partial L\over \partial q^i}(t,c(t),\dot{c}(t)).
\]
For all fixed time $t\in [t_0,t_1]$, the Legendre transform is the mapping
\[
\begin{array}{cccc}
T{\cal Y} & \longrightarrow & T^*{\cal Y}\\
(q,v) & \longmapsto & (q,p)=(q,\partial L(t,q,v)/\partial v),
\end{array}
\]
where $q\in {\cal Y}$, $v\in T_q{\cal Y}$ and $p\in T^*_q{\cal Y}$. In cases
where, for all time $t$, this mapping is a diffeomorphism, we define the
Hamiltonian function $H:[t_0,t_1]\times T^*{\cal Y}\longrightarrow \Bbb{R}$ by
\[
H(t,q,p) := \langle p,V(t,q,p)\rangle - L(t,q,V(t,q,p)),
\]
where $(q,p)\longmapsto (q,V(t,q,p))$ is the inverse mapping of the Legendre mapping.
Then it is well-known that $t\longmapsto c(t)$ is a solution of the Euler--Lagrange
equations if and only if $t\longmapsto (c(t),\partial L(t,c(t),\dot{c}(t))/\partial v)=:(c(t),\pi(t))$
is a solution of the {\em Hamilton equations}
\[
{dc^i\over dt}(t) = {\partial H\over \partial p_i}(t,c(t),\pi(t)),\quad \hbox{and}\quad
{d\pi_i\over dt}(t) = -{\partial H\over \partial q^i}(t,c(t),\pi(t)).
\]
Thus this converts the second order Euler--Lagrange equations into the flow equation of the
non autonomous vector field $X_{H,t}$ defined over $T^*{\cal Y}$ by
\begin{equation}\label{1.hamilton}
X_{H,t}\iN \Omega + dH_t = 0.
\end{equation}
where $\Omega := \sum_{i=1}^kdp_i\wedge dq^i$ is the symplectic form over $T^*{\cal Y}$,
$H_t(q,p):= H(t,q,p)$ and ``$\iN $''  denote the interior product, i.e.\,for any tangent vector
$\xi\in T_{(q,p)}\left(T^*{\cal Y}\right)$,
$\xi\iN \Omega$ is the 1-form such that $\xi\iN \Omega(V) = \Omega (\xi,V)$,
$\forall V\in T_{(q,p)}\left(T^*{\cal Y}\right)$\\

\noindent
Instead of viewing the dynamics as the motion of a point in some space, like
for instance the phase space $T^*{\cal Y}$, we can use another approache which consists in determining how an observable
quantity, such as the position or the momentum of a particle, evolves. This is achieved by
using the Poisson bracket operation
\[
\begin{array}{ccc}
{\cal C}^\infty (T^*{\cal Y})\times {\cal C}^\infty (T^*{\cal Y}) &
\longrightarrow & {\cal C}^\infty (T^*{\cal Y})\\
(F,G) & \longmapsto & \{F,G\},
\end{array}
\]
where
\[
\{F,G\} := \sum_{i=1}^k\left( {\partial F\over \partial p_i}{\partial G\over \partial q^i}
- {\partial F\over \partial q^i}{\partial G\over \partial p_i}\right) .
\]
Then, for all Hamiltonian trajectory $t\longmapsto(c(t),\pi(t))$, and for all
$F\in {\cal C}^\infty (T^*{\cal Y})$, we have
\[
{dF(c(t),\pi(t))\over dt} = \{H,F\}(c(t),\pi(t)).
\]
For example in the particular case where the variational problem is autonomous
(i.e.\,$L$ does not depend on $t$) then $H$ does not depend on time and we deduce
from the skewsymmetry of the Poisson bracket
that the energy is conserved along the trajectories, a special case
of Noether theorem when the problem is invariant by time translation.\\

\noindent
Eventually this formulation of the dynamics is a good preliminary for modelling
the evolution of the quantum version of our problem: for instance by replacing the functions in
${\cal C}^\infty (T^*{\cal Y})$ by Hermitian self-adjoint operators and the
Poisson bracket by the commutator $[\cdot ,\cdot]$
we ``guess'' the Heisenberg evolution equation 
\[
i\hbar {d\widehat{F}\over dt} = [\widehat{H},\widehat{F}],
\]
consequently the commutation relations $[\widehat{p}_i,\widehat{q}^j] = i\hbar \delta^j_i$, is nothing but a
formalisation of Heisenberg incertainty principle.\\

\noindent
All that leads to a now ``well understood'' strategy of building a
mathematical description of a quantum particle governed by a Hamiltonian functions $H$
(with the restriction that, among other things, the correspondence
$H\longmapsto \widehat{H}$ is far from being uniquely defined).
Starting from a variational formulation
of Newton's law of mechanics, it is thus possible to formally derive the Schr\"odinger
equation more or less by following the steps discussed above.\\

\noindent
Now the more challenging question is
to produce a similar analysis for fields theories.\\

\noindent
Quantum fields theory\footnote{Starting from the frame­work of classical physics, the concept of a field at first might
evoke ideas about macroscopic systems, for example velocity fields or temperature fields in fluids and gases, etc. Fields
of this kind will not concern us, however; they can be viewed as derived quantities which arise from an averaging of
microscopic par­ticle densities. Our subjects are the fundamental fields that describe matter on a microscopic level:
it is the quantum-mechanical wave function $\psi(x, t)$ of a system which can be viewed as a field from which the
observable quantities can be deduced. In quantum mechanics the wave function is introduced as an ordinary complex-valued
function of space and time. In Dirac's terminology it has the character of a ``{\em c} number''. Quantum field theory goes
one step further and treats the wave function itself as an object which has to undergo quantization. In this way the wave
function $\psi(x, t)$ is transmuted into a field operator $\hat{\psi}(x,t)$, which is an operator-valued quantity (a
``{\em q} number'') satisfying certain commutation relations. This process, often called ``second quantization'', is quite
analogous to the route that in ordinary quantum mechanics leads from a set of classical coordinates $q_i$ to a set of
quantum operators $\hat{q}_i$. There is one important technical difference, though, since $\hat{\psi}(x, t)$ is a field,
i.e.\,an object which depends on the coordinate $x$. The latter plays the role of a ``continuous-valued'' index, in contrast
to the discrete index $i$, which labels the set $q_i$. Field theory therefore is concerned with systems having an infinite
number of degrees of freedom.
The concept of field quantization has far-reaching consequences and is one of the cornerstones of modern physics.
Field quantization provides an elegant language to describe particle systems. Moreover the theory naturally leads to
the insight that there are field quanta which can be created and annihilated. These field quanta come in many guises
and are found in virtually all areas of physics.} results from the efforts of physicists in order to cure the shortcoming
of the Schr\"odinger equation. Indeed, this latter equation is not invariant by the
group of special relativity, the Poincar\'e group, but by the (projective) Galilean group.
This was one of the motivations which led Dirac to its famous equation.
Another concern was the interactions of charged particles and a relativistic field, namely the electro-magnetic field
governed by Maxwell equations.
A second reason for fields theory is that, unlike the Schr\"odinger equation, they
allow interactions of a variable number of particles. The simplest exemple is
the Klein--Gordon equation for scalar fields $\varphi:\Bbb{R}^{1,3}\longrightarrow \Bbb{R}$ (or $\Bbb{C}$):
\[
{1\over c^2}{\partial ^2\varphi\over \partial t^2} - \Delta \varphi +m^2\varphi = 0.
\]
This is the Euler--Lagrange equation for the  variational problem
\[
{\cal L}[\varphi]:= \int_{\Bbb{R}^{1,3}}{1\over 2}\left(
{1\over c^2}\left|{\partial \varphi\over \partial t}\right|^2 -
|\nabla \varphi |^2 - m^2|\varphi|^2\right)dtdx^1dx^2dx^3.
\]
Here $\Bbb{R}^{1,3}$ is the Minkowski space. Note that the integral
${\cal L}[\varphi]$ may not be defined, but if $\varphi$ smooth or in $H^1_{loc}$ then,
for any compact subset $K\subset \Bbb{R}^{1,3}$, the integral ${\cal L}_K[\varphi]$ of the Lagrangian
density over $K$ is defined  and we say that $\varphi$ is a critical point of ${\cal L}$
if and only if, for any $K$ the restriction of $\varphi$ to $K$ is a critical point of
${\cal L}_K$. At this level we can address the following questions: is it possible to follow the same lines as for a
1-dimensional variational problem and  build a Hamiltonian formulation of the
Klein--Gordon equation ?
And can we deduce a quantization procedure for that equation ?\\

\noindent
The answer is positive using an approache developed by physicists and is known
as the canonical quantization of fields: one chooses a time coordinate $t$ over the Minkowski space-time
and  looks, for any value of $t_0$, at the instantaneous state of the field, i.e.\,the
restriction of $\varphi$ on the Cauchy hypersurface $\{t=t_0\}$. Then we picture the whole
history of the field as an evolution, parametrized by $t$, of a point in the infinite
dimensional space $\{(x^1,x^2,x^3)\longmapsto \varphi (x^1,x^2,x^3)\}$. One associates to each
path in this infinite dimensional space an action, which is just the one obtained by using
Fubini theorem through the splitting $\Bbb{R}^{1,3}\simeq \Bbb{R}\times \Bbb{R}^3$.
Then we follow the same procedure as the one that we described at the begining of this paper,
but this time in some infinite dimensional manifold: we consider the cotangent bundle to
the manifold $\{(x^1,x^2,x^3)\longmapsto \varphi (x^1,x^2,x^3)\}$ and, using
the Legendre transform obtained from the Lagrangian, we build a conjugate (momentum)
variable $[(x^1,x^2,x^3)\longmapsto \pi(t,x^1,x^2,x^3)]$ (which in this case is just $[(x^1,x^2,x^3)\longmapsto
{\partial \varphi\over \partial t}(t,x^1,x^2,x^3)]$).
Then one can write Hamilton equations, Poisson bracket and deduce the canonical quantization.\\

\noindent
Note that all that is relatively formal but, after each step, it is possible to formulate
a theory which makes sense mathematically. This is because of the very particular
structure of the Klein--Gordon equation, being hyperbolic and linear. Similarly
one can perform a quantization of Maxwell equations (with some extra work due to the
gauge invariance). But everything breaks down as soon as the equation is nonlinear,
even if the nonlinearity is very mild (the best that we can do is to compute
physically relevant quantities by perturbation, even if there is no mathematical bases).
So one can quantize only extremely particular variational problems. 
Beside this outstanding difficulty we are faced with a further critic which
is that all this procedure does not respect relativistic invariance: indeed, we were obliged to
choose a time coordinate from the begining in order to perform a Legendre transform
and then to write down Hamilton's equations, and so on. Commonly, we say that this theory is not
{\em covariant}, i.e.\,it does not respect the (general) relativity principle which implies the independence
toward a used coordinate or reference system. Consequently, this description is not quite satisfactory, even
if actually one can check that the resulting quantum field does not depend on the time
coordinate which was used.\\

\noindent
We want here to consider seriously this critic:
is it possible to follow a more covariant path ? A method is well-known:
it is the Feynman integral approach, based on the Lagrangian formulation, without
using the Hamiltonian framework. It is much more suitable than the canonical approach
for computing ``correlations'' for nonlinear (i.e.\,interacting) theories. However
it is less constructive than the canonical approach, which has the advantage of providing us
with a scheme to construct the Hilbert Fock space and the operators.
So is there a covariant Hamiltonian approach ? In principle it should be possible and
this was suggested independentely by M. Born \cite{Born} and H. Weyl \cite{Weyl} in 1935:
it would be based on using a covariant Hamiltonian
formalism for variational problems with several variables quite different from the one
that we described above, which used a slicing of space-time.\\

\noindent
The first example of such a formalism was built by C. Carath\'eodory \cite{Cara}
and another version was proposed later independentely by H. Weyl \cite{Weyl} and T. de Donder
\cite{dDW}. Note that in constrast with the 1-dimensional calculus of variations,
there are actually infinitely many possible theories. There were described by T.H.J.
Lepage \cite{Lepage} and H. Boerner \cite{Boerner}. We shall see later how to understand
within a global picture this multiplicity of formalisms. Before that we will
expound the de Donder--Weyl theory, since it is the simplest one.

\section{The de Donder--Weyl theory}

The historical background of the formalism expounded in this section
is deeply rooted in the work of C. Carath\'eodory, H. Weyl and T. de Donder,
followed by the observation by E. Cartan \cite{Cartan}
(who called $\theta^{dDW}$ the {\em de Donder form}) in 1933.
But it seems really
to have been developped under the impulsion of W. Tulczyjew \cite{Tulczyjew1} in 1968
and the Polish school of mathematical physics: J. \'{S}niatycki \cite{Snyatycki},
K. Gaw\c{e}dski \cite{Gawedski}, J. Kijowski \cite{Kijowski1}, 
J. Kijowski and W. Szczyrba \cite{KijowskiSzczyrba}, J. Kijowski and W.M. Tulczyjew
\cite{KijowskiTulczyjew}, and through the papers of
P.L. Garc\'\i a and A. P\'erez-Rend\'on \cite{GarciaPerez} and
H. Goldschmidt and S. Sternberg \cite{GoldschmidtSternberg}.
Various descriptions and points of view about the foundations of these theories
can also be found in the books \cite{Binz}, \cite{GiaquintaHildebrandt}, \cite{Rund}
and \cite{Sardanashvily} or in the papers \cite{Gotay1}, \cite{Gotay2}, \cite{Gotay3},
\cite{Kastrup}, \cite{Martin}.\\

\noindent
Let us consider the 4-dimensional space-time $\Bbb{R}^{1,3}$ as a source domain and
$\Bbb{R}^k$ as a target space. A first order action functional on maps $u:\Bbb{R}^{1,3}\longrightarrow \Bbb{R}^k$ is defined by means of
a Lagrangian density $L$: it is a function on the variables
$(x,y,v)$, where $x\in \Bbb{R}^{1,3}$, $y\in \Bbb{R}^k$ and
$v\in \left(\Bbb{R}^{1,3}\right)^*\otimes \Bbb{R}^k$ is a linear map
from $\Bbb{R}^{1,3}$ to $\Bbb{R}^k$ (alternatively we may consider $v$ as a $n\times k$ real matrix).
Then we define the functional
\[
{\cal L}[u]:= \int_{\Bbb{R}^{1,3}}L(x,u(x),du(x))\omega,
\]
where $\omega:= dx^0\wedge dx^1\wedge dx^2\wedge dx^3$. The Euler--Lagrange equation for
critical points is
\[
{\partial \over \partial x^\mu}\left( {\partial L\over \partial v_\mu^i}(x,u(x),du(x))\right)
= {\partial L\over \partial y^i}(x,u(x),du(x)).
\]
The de Donder--Weyl theory is simply based on the change of variables
\[
\pi^\mu_i(x):= {\partial L\over \partial v_\mu^i}(x,u(x),du(x)),
\]
i.e.\,exchanging the variables $(x,y,v)$ with $(x,y,p)$ where
$p\in \Bbb{R}^{1,3}\otimes \left(\Bbb{R}^k\right)^*$
is given by $p^\mu_i=\partial L/\partial v_\mu^i(x,y,v)$. This
works of course if we make the assumption that $(x,y,v)\longmapsto (x,y,p)$
is a diffeomorphism, an analog of the Legendre condition. In the following we shall
suppose that this Legendre hypothesis is satisfied (but very interesting
situations occur when precisely this Legendre condition fails, as for instance
in the case of gauge theories). Then we can consider the Hamiltonian function
on $\Bbb{R}^{1,3}\times \Bbb{R}^k\times\left(\Bbb{R}^{1,3}\otimes \left(\Bbb{R}^k\right)^*\right)$
\[
H(x,y,p):= p^\mu_iv_\mu^i - L(x,y,v),\quad \hbox{where }v\hbox{ is defined
implicitely by}
\quad
p^\mu_i={\partial L\over \partial v_\mu^i}(x,y,v).
\]
A simple computation shows that the Euler--Lagrange equations are
equivalent to the system of generalized Hamilton equations
\begin{equation}\label{2.hamilton/brut}
\sum_\mu{\partial \pi^\mu_i(x)\over \partial x^\mu} =
- {\partial H\over \partial y^i}(x,u(x),\pi(x)),
\quad 
{\partial u^i(x)\over \partial x^\mu} =
{\partial H\over \partial p^\mu_i}(x,u(x),\pi(x)).
\end{equation}
This is a simple example of a more general situation.
We can replace for example $\Bbb{R}^{1,3}$ by 
a smooth $n$-dimensional oriented manifold ${\cal X}$ 
and $\Bbb{R}^k$ by another $k$-dimensional manifold ${\cal Y}$.
Then the Lagrangian density $L$ is a function on variables $x\in {\cal X}$,
$y\in {\cal Y}$ and $v\in T_y{\cal Y}\otimes T^*_x{\cal X}$. 
Thus $L$ can be seen as a smooth function defined on the bundle over ${\cal X}\times
{\cal Y}$ with fiber over $(x,y)$ equal to $T_y{\cal Y}\otimes T^*_x{\cal X}$.
We denote by $T{\cal Y}\otimes _{{\cal X}\times {\cal Y}}T^*{\cal X}$ this bundle.
Using some volume $n$-form $\omega$ on ${\cal X}$ we hence define the functional
${\cal L}[u]:= \int_{\cal X}L(x,u(x),du(x))\omega$
on the set of maps $u:{\cal X}\longrightarrow {\cal Y}$. A similar Legendre transform
can be defined, leading to a Hamiltonian function defined on the {\em multisymplectic
manifold} $T^*{\cal Y}\otimes _{{\cal X}\times {\cal Y}}T{\cal X}$.\\

\noindent
We shall see now that this manipulation has some geometrical
interpretation in a way analogous to the 1-dimensional calculus of
variations. The more naive way to formulate it
consists to associate to any pair of maps
$x\longmapsto (u(x),\pi(x))$ its graph $\Gamma$ in
$T^*{\cal Y}\otimes _{{\cal X}\times {\cal Y}}T{\cal X}$, i.e.\,the
image of ${\cal X}\ni x\longmapsto (x,u(x),\pi(x))$. $\Gamma$ is the
$n$-dimensional analog of a curve in a symplectic manifold. Then
at each point $(x,u(x),\pi(x))$ of $\Gamma$ we attach the tangent
$n$-multivector $X\in \Lambda^nT_{(x,u(x),\pi(x))}
\left( T^*{\cal Y}\otimes _{{\cal X}\times {\cal Y}}T{\cal X}\right)$
defined by
\[
X:= X_1\wedge \cdots \wedge X_n,\quad \hbox{where}\quad
X_\mu:= {\vec{\partial \over \partial x^\mu}} + {\partial u^i(x)\over \partial x^\mu}
{\vec{\partial \over \partial y^i}} + {\partial \pi^\nu_i(x)\over \partial x^\mu}
{\vec{\partial \over \partial p^\nu_i}}.
\]
Now we define on the multisymplectic manifold
$T^*{\cal Y}\otimes _{{\cal X}\times {\cal Y}}T{\cal X}$
the {\em multisymplectic $(n+1)$-form}
\[
\Omega^*:= \sum_\mu \sum_i dp^\mu_i\wedge dy^i\wedge \omega_\mu,
\]
where $\omega_\mu:= {\vec\partial} /\partial x^\mu\iN \omega$, i.e.\,$\omega_\mu$ is the unique
$(n+1)$-form such that $\forall V_1,\cdots ,V_{n-1}\in
T_{(x,y,p)}\left( T^*{\cal Y}\otimes _{{\cal X}\times {\cal Y}}T{\cal X}\right)$,
$\omega_\mu(V_1,\cdots ,V_{n-1})= \omega({\vec\partial} /\partial x^\mu,V_1,\cdots ,V_{n-1})$.\\

\noindent
We also define the interior product of $X$ by $\Omega^*$ to be the
unique 1-form $X\iN \Omega^*$ such that
$\forall V\in T_{(x,y,p)}\left( T^*{\cal Y}\otimes _{{\cal X}\times {\cal Y}}T{\cal X}\right)$,
$X\iN \Omega^*(V)= \Omega^*(X_1,\cdots ,X_n,V)$. Then we can compute that
\[
X\iN \Omega^* = (-1)^n\left( {\partial u^i(x)\over \partial x^\mu}dp^\mu_i
- {\partial \pi^\mu_i(x)\over \partial x^\mu}dy^i\right).
\]
Hence, by comparing this last expression with 
the value of $dH = {\partial H\over \partial x^\mu}dx^\mu + 
{\partial H\over \partial y^i}dy^i + {\partial H\over \partial p^\mu_i}dp^\mu_i$
at $(x,u(x),\pi(x))$, we see that $\Gamma$ is the graph of a solution of
the Hamilton system of equations (\ref{2.hamilton/brut}) if and only if
\begin{equation}\label{2.hamilton/moyen}
X\iN \Omega^* = (-1)^n dH_{(x,u(x),\pi(x))}\quad \hbox{mod}\quad
dx^\mu.
\end {equation}
Here ``mod $dx^\mu$'' means that the equality holds between the coefficients
of $dy^i$ and $dp^\mu_i$ in both sides. This looks like the analog
of the classical Hamilton equation (\ref{1.hamilton}) (except that here we do have the
``mod $dx^\mu$'' restriction). This suggests us to use $\Omega^*$ as a replacement for the
standard symplectic form. Note that $\Omega^*$ is the differential of the
generalized {\em Poincar\'e--Cartan $n$-form} $\theta^*:= \sum_\mu \sum_i p^\mu_i dy^i\wedge \omega_\mu$.
Moreover, as in the case
of classical mechanics, the generalized Poincar\'e--Cartan form encodes
all the informations that we need in order to perform the Legendre transform.\\

\noindent
Let us discuss now the second question --- about the Poisson bracket. It has to be
defined on functionals on the space of solutions to the generalized Hamilton
equations. Thus we first consider the space ${\cal E}$ of all smooth $n$-dimensional
submanifolds $\Gamma$ of the multisymplectic manifold
which are graphs of the generalised Hamilton equations, i.e.\,such
that for any $m\in \Gamma$, there exists an $n$-multivector $X$ which is tangent to $\Gamma$
at $m$ and which satisfies equation (\ref{2.hamilton/moyen}). We call a {\em Hamiltonian
$n$-curve} any such submanifold. Among the functionals ${\cal E}\longrightarrow \Bbb{R}$
we shall restrict ourself to a particular class: this choice is motivated by the
particular observable quantities used by physicists in quantum field theory. Indeed
the observable functionals in the canonical field theory are integrals (smeared
with smooth test
functions) over a spacelike hypersurface of the space-time of either the values of fields
components or the value of their time first derivative (in the latter case the time
derivative appear because we are actually considering momenta). And it turns out that
(almost) all such observable functionals (see the next section) can be described by a pair $(\Sigma,F)$, where
$\Sigma$ is a codimension 1 submanifold of the multisymplectic manifold and
$F$ is a $(n-1)$-differential form on the multisymplectic manifold such that
there exists a tangent vector field $\xi_F$ such that
\[
dF + \xi_F\iN \Omega^* = 0.
\]
$F$ is then called an {\em algebraic observable} $(n-1)$-form.
Then, if $\Sigma$ satisfies suitable transversality conditions (see \cite{HeleinKouneiher2}),
$\Sigma\cap \Gamma$ is a $(n-1)$-dimensional manifold and we define the functional
$\int_\Sigma F$ to be
\[
\begin{array}{ccc}
{\cal E} & \longrightarrow & \Bbb{R}\\
\Gamma & \longmapsto & \int_{\Sigma \cap \Gamma} F.
\end{array}
\]
Then a bracket can be defined between two observable functionals $\int_\Sigma F$ 
and $\int_\Sigma G$ by
\[
\left\{ \int_\Sigma F,\int_\Sigma G\right\} := \int_\Sigma \{F,G\},
\quad \hbox{where}\quad \{F,G\}:= \xi_F\wedge \xi_G\iN \Omega^*.
\]
Here $\xi_F\wedge \xi_G\iN \Omega^*$ is the unique $(n-1)$-form such that
for all tangent vectors $V_1,\cdots ,V_{n-1}$,
$\xi_F\wedge \xi_G\iN \Omega^*(V_1,\cdots ,V_{n-1}) =
\Omega^*(\xi_F,\xi_G,V_1,\cdots ,V_{n-1})$. It can be checked that this Poisson bracket
satisfies ``good'' properties: first of all it coincides with the standard Poisson
bracket ot the canonical theory of fields, second it satisfies the Jacobi identity,
if for instance the observable $(n-1)$-forms decrease to zero at
infinity\footnote{nevertheless a notable difference with the Poisson bracket
of classical mechanics is that we cannot easily make sense of the product of two such
forms. Attempts in this direction are proposed in \cite{Kanatchikov1}; we suggest an alternative
point of view at the end of Paragraph 5.4 in this text}.\\

\noindent
The relatively naive description presented here deserves some critics: 
\begin{itemize}
\item Equation (\ref{2.hamilton/moyen}) holds only ``mod $dx^\mu$'', which is not
very aesthetic: this reflects a disymmetry between the space-time variables
and the field component variables.
\item Very important ``observable'' quantities are the components of the energy-momentum
tensor $T_{\mu\nu}$. This tensor is linked through the Noether theorem to space-time
translations symmetries in special relativity or to diffeomorphism invariance
in general relativity. It is important in quantum field theory, where it helps to construct
the Hamiltonian in the standard canonical picture, but also crucial in general relativity,
since it models the way the distribution of energy and matter bend the space-time
geometry through the Einstein equation $R_{\mu\nu} - {1\over 2}Rg_{\mu\nu} = T_{\mu\nu}$.
In the previous setting there is no clear representation of the energy-momentum
tensor.
\item Most important variational problems in physics involve fiber bundles:
the Maxwell--Dirac, the Yang--Mills--Dirac or the Yang--Mills--Higgs theories for
the electrodynamics, the electroweak and the strong forces and also the general relativity.
They does not fit into the above formalism:
indeed the field components cannot be treated as coordinates on an independant
manifold and one needs a more general framework.
\end{itemize}
The two first critics can be cured by adding to the set of variable $(x,y,p)$
a further variable $e\in \Bbb{R}$, canonically conjugate to the space-time
volume form $\omega$. Then we consider on
$\left( T^*{\cal Y}\otimes _{{\cal X}\times
{\cal Y}}T{\cal X}\right)\times \Bbb{R}$ the multisymplectic form
\[
\Omega^{dDW}:= de\wedge \omega + \sum_\mu \sum_i dp^\mu_i\wedge dy^i\wedge \omega_\mu
= de\wedge \omega + \Omega^*.
\]
Any solution of the Hamilton equations can be represented by a smooth $n$-dimensional
submanifold $\Gamma$ in
$\left( T^*{\cal Y}\otimes _{{\cal X}\times {\cal Y}}T{\cal X}\right)\times \Bbb{R}$
such that, instead of equation (\ref{2.hamilton/moyen}) we have
\begin{equation}\label{2.hamilton}
X\iN \Omega^{dDW} = (-1)^n d{\cal H}_{(x,u(x),\epsilon(x),\pi(x))},\quad
\hbox{where}\quad
{\cal H}(x,y,e,p):= e + H(x,y,p).
\end{equation}
Indeed it can be achieved by choosing $\epsilon(x)$ such that
${\cal H}(x,u(x),\epsilon(x),\pi(x))$ is constant along $\Gamma$.
We can hence avoid the ``mod $dx^\mu$''
restriction. Moreover $\Omega^{dDW}$ is the differential of a Poincar\'e--Cartan form
$\theta^{dDW}:= e\, \omega + p^\mu_idy^i\wedge \omega_\mu$ and the component of the stress-energy
tensor can be recovered as coefficients of the observable $(n-1)$-forms
${\partial \over \partial x^\mu} \iN \theta^{dDW}$.\\

\noindent
Thus it remains to understand better the third point: to find a geometrical framework
for generalizing the above construction to more general variational problems.
This question is even more accurate now since we added
a further variable $e$ whose geometrical sense needs to be understood. For all variational
problems on fields which are sections of a bundle ${\cal F}$, the usual approach is to consider
the affine first jet bundle associated to these sections and to build the multisymplectic
manifold as a dual of this affine jet bundle. References concerning this approach
are \cite{gimmsy} or \cite{Echeverria}.

\section{A general framework: multisymplectic manifolds}
The above theory is an example of a multisymplectic manifold $({\cal M},\Omega)$: a differential manifold
${\cal M}$ equipped with a multisymplectic $(n+1)$-form $\Omega$. An $(n+1)$-form
$\Omega$ is {\em multisymplectic} if and only if
\begin{itemize}
\item it is closed: $d\Omega = 0$
\item it is nondegenerate: $\forall \xi\in T_m{\cal M}$,
$\xi\iN \Omega = 0\ \Longrightarrow \ \xi = 0$.
\end{itemize}
Given a multisymplectic manifold $({\cal M},\Omega)$ we can define the notion of
algebraic observable $(n-1)$-forms and use them,
together with hypersurfaces $\Sigma$ to define observable functionals
as in the preceeding section. Poisson brackets are defined in a similar way.\\

\noindent
There is also a notion which leads to a generalization of the standard relation
${dF\over dt}=\{H,F\}$ of classical mechanics.
Given the Hamiltonian function ${\cal H}$, for any algebraic observable $(n-1)$-form $F$
we let
\begin{equation}\label{pseudobracket}
\{{\cal H}, F\}:= - d{\cal H}(\xi_F).
\end{equation}
Although this definition and the notation used here are reminiscent from those
of Poisson bracket, it is not clear whether we may consider this operation a Poisson
bracket (because in particular, ${\cal H}$ is not an observable form nor any observable
quantity).
Thus we prefer to call this operation a Poisson {\em pseudobracket}.
Anyway it is related to the following useful result: assume that $\Gamma$ is a Hamiltonian
$n$-curve and $F$ is an algebraic observable $(n-1)$-form, then we have
\begin{equation}\label{2.dH}
dF_{|\Gamma} = \{{\cal H},F\}\omega_{|\Gamma},
\end{equation}
where $dF_{|\Gamma}$ is the restriction of $dF$ to $\Gamma$
(see \cite{Kanatchikov1}, \cite{HeleinKouneiher1}). The proof is straightforward:
for all $q\in \Gamma$ we let $(X_1,\cdots ,X_n)$ be a basis of $T_q\Gamma$ such that
$\omega(X_1,\cdots ,X_n)=1$. Then
$dF(X_1,\cdots ,X_n) = -\xi_F\iN \Omega(X_1,\cdots ,X_n)
= -(-1)^nX_1\wedge \cdots \wedge X_n\iN \Omega(\xi_F)
= -d{\cal H}(\xi_F) = \{{\cal H},F\}$.\\

\noindent
However
in a work in collaboration with J. Kouneiher in \cite{HeleinKouneiher2} we point out that
the class of algebraic observable $(n-1)$-forms can be enlarged as follows: an $(n-1)$-form $F$ is
called an {\em observable} $(n-1)$-form if and only if, $\forall q\in {\cal M}$,
if $X$ and $\widetilde{X}$ are two decomposable $n$-multivectors in $\Lambda^nT_q{\cal M}$
such that
$X\iN \Omega = \widetilde{X}\iN \Omega$, then $dF(X)=dF(\widetilde{X})$ (see
\cite{HeleinKouneiher2} for details).
One can then define the pseudobracket ${\{\cal H},F\}$ to be equal to $dF(X)$, where
$X$ is any decomposable $n$-multivector such that $X\iN \Omega = (-1)^nd{\cal H}$.
It is easy to show that this pseudobracket agrees with the previous one (\ref{pseudobracket})
for algebraic observable $(n-1)$-forms and that
the dynamical relation (\ref{2.dH}) can hence be generalized
to (non necessarily algebraic) observable $(n-1)$-forms. This definition 
and the underlying point of view differ
from the one previously used in the litterature (which corresponds to algebraic
observable $(n-1)$-forms). We believe that this point of view is more natural,
being directly related to the fundamental identity (\ref{2.dH}). When we shall follow
this point of view in the next Section to the question of observable
$(p-1)$-forms, for $1\leq p<n$, it will lead also to a new definition of these
observable $(p-1)$-forms.\\

\noindent
Now what is the difference between the
two definitions ? On the one hand every algebraic observable $(n-1)$-form $F$ is an observable
$(n-1)$-form since $X\iN \Omega = \widetilde{X}\iN \Omega$ implies
$dF(X) = -\xi_F\iN \Omega(X) = -\xi_F\iN \Omega(\widetilde{X}) = dF(\widetilde{X})$.
On the other hand the converse is false: consider for example the de Donder--Weyl
theory expounded above for maps $u:\Bbb{R}^n\longrightarrow \Bbb{R}^k$, for $k>1$.
Then $y^1dy^2\wedge dx^3\wedge \cdots \wedge dx^n$ is observable but not algebraic
observable. We propose to call a multisymplectic manifold on which the set of
algebraic observable $(n-1)$-forms coincide with the set of observable $(n-1)$-forms
a {\em pataplectic manifold}. We shall see examples in the next section.

\section{Generalizations}
The formalism that we have presented in Section 2 is based on the de Donder--Weyl theory,
which is a particular case among infinitely many theories which were classified
in 1936 by T.H.J. Lepage \cite{Lepage}. A geometric and universal setting for describing
simultaneously all these theories was expounded first by P. Dedecker in 1953
\cite{Dedecker}. 
Here the idea is that we can view each first order variational problem as
a variational problem on a class of $n$-dimensional submanifolds $G$ of some
manifold ${\cal N}$ of dimension $n+k$:
$G$ could be the graph of a map between two manifolds, the
image of a section of a fiber bundle or something more general. Then
the Lagrangian density can be identified with a function $L$ defined
on the Grassmannian bundle $Gr^n{\cal N}$, i.e.\,the bundle over ${\cal N}$ whose
fiber at any point $q\in {\cal N}$ is the set of oriented $n$-dimensional
vector subspaces of $T_q{\cal N}$. This is a kind of analog of the (projective)
tangent bundle of a manifold that is used in Lagrangian Mechanics. Then the
analog of the cotangent bundle is the bundle of differential $n$-forms on ${\cal N}$,
$\Lambda^nT^*{\cal N}$. Note that, in contrast with classical mechanics (which corresponds
to $n=1$) or with the de Donder--Weyl theory, $1+\hbox{dim}Gr^n{\cal N} = 1+n+k+nk$ is
in general strictly less than $\hbox{dim}\Lambda^nT^*{\cal N} = n+k+{(n+k)!\over n!k!}$.
So the Legendre transform is replaced by a Legendre correspondence which --- generically ---
associates to each ``multivelocity'' $T\in Gr^n_q{\cal N}$ an affine subspace
of $\Lambda^nT^*_q{\cal N}$ of dimension ${(n+k)!\over n!k!}-nk-1$ called
{\em pseudofiber} by Dedecker. Now each Lepage theory (one instance being the de Donder--Weyl
one, when it makes sense) corresponds to choosing a submanifold of $\Lambda^nT^*{\cal N}$
which intersects transversally all pseudofibers through exactly one point (if the
Legendre condition holds).\\

\noindent
Note that, to my knowledge, almost all the literature on the subject focuses on the de Donder--Weyl theory,
excepted \cite{Dedecker} and \cite{Kijowski2}. It seems however important 
(if in particular we are interested in gravitation theories) to understand
all the development expounded in the previous paragraph in the Lepage--Dedecker framework.
This has been addressed in collaboration with J. Kouneiher in our papers \cite{HeleinKouneiher1}
and \cite{HeleinKouneiher2}.\\

\noindent
In order to understand the difference let us see a very simple example: variational
problems on maps $u:\Bbb{R}^2\longrightarrow \Bbb{R}^2$. Let us denote by
$\omega:=dx^1\wedge dx^2$ the volume form on the domain space $\Bbb{R}^2$.
A map is pictured by its graph,
a 2-dimensional submanifold $G$ of $\Bbb{R}^4=\Bbb{R}^2\times \Bbb{R}^2$, whose
projection on the first factor $\Bbb{R}^2$ is a diffeomorphism (or equivalentely
s.t.\,$\omega_{|G}\neq 0$). Given a point $(x,y)\in G\subset \Bbb{R}^4$ the
tangent plane to $G$ at $(x,y)$ is spanned by vectors
$X_1= {\vec{\partial \over \partial x^1}}+v_1^i{\vec{\partial \over \partial y^i}}$ and
$X_2={\vec{\partial \over \partial x^2}}+v_2^i{\vec{\partial \over \partial y^i}}$,
where $(v^1_1,v^1_2,v^2_1,v^2_2)\in \Bbb{R}^4$. So the set of all possible
tangent planes to such $G$'s is parametrized by the variables $v^i_\mu$ and we can use
the local coordinates $x^\mu$, $y^i$ and $v^i_\mu$ on $Gr^2\Bbb{R}^4$.
Now the analog of the cotangent bundle in this situation is $\Lambda^2T^*{\cal N}$.
Using coordinates $x^\mu$ and $y^i$ on $\Bbb{R}^4$, a basis of the 6-dimensional space
$\Lambda^2T^*_{(x,y)}\Bbb{R}^4$ is $(dx^1\wedge dx^2,dy^i\wedge dx^\mu,dy^1\wedge dy^2)$.
Thus any 2-form $P\in \Lambda^2T^*_{(x,y)}\Bbb{R}^4$ can be identified with the coordinates $(e,p^\mu_i,r)$ such that $P = 
e\,dx^1\wedge dx^2 + \epsilon_{\mu\nu}p^\mu_idy^i\wedge dx^\nu + r\,dy^1\wedge dy^2$,
where $\epsilon_{12}=-\epsilon_{21}=1$ and $\epsilon_{11}=\epsilon_{22}=0$.\\

\noindent
Now the Legendre correspondence is obtained by the following.
Given some $P\simeq (e,p^\mu_i,r)\in \Lambda^2T^*_{(x,y)}\Bbb{R}^4$ and some tangent
space $T\simeq (v^i_\mu)\in Gr^2_{(x,y)}{\cal N}$, we define the pairing
$\langle T,P\rangle:=P(X_1,X_2)$, where $(X_1,X_2)$ forms a basis of $T$ such
that $\omega(X_1,X_2)=1$. Using local coordinates we have here
$\langle T,P\rangle = e+p^\mu_iv^i_\mu+r(v^1_1v^2_2-v^1_2v^2_1)$. We also
define the function $W(x,y,T,P) = \langle T,P\rangle - L(x,y,T)$, where $L$ is the
Lagrangian density (identified here with a function on $Gr^2\Bbb{R}^4$). Then we say
that $T$ is in correspondence with $P$ if and only if ${\partial W\over \partial T}(x,y,T,P)=0$.
This relation writes in local coordinates
\[
p_i^\mu + \epsilon^{\mu\nu}\epsilon_{ij}v_\nu^j\,r =
{\partial L\over \partial v^i_\mu}(x^\mu,y^i,v^i_\mu).
\]
As announced previously, given some $(x^\mu,y^i,v^i_\mu)$ the solution
to this equation is in general not unique, but it is actually an affine plane
(the pseudofiber)
inside $\Lambda^2T^*_{(x,y)}\Bbb{R}^4$, parallel to the vector plane spanned by
\[
dx^1\wedge dx^2\quad \hbox{and}\quad
\left( v_1^1v_2^2 -v_1^2v_2^1 \right) dx^1\wedge dx^2 -
\epsilon_{ij}v_\nu^j dy^i\wedge dx^\nu + dy^1\wedge dy^2.
\]
Hence, inside $\Lambda^2T^*_{(x,y)}\Bbb{R}^4$,
pseudofibers form a 4-parameters family of non parallel affine planes. The subset 
${\cal M}_{(x,y)}$ of $\Lambda^2T^*_{(x,y)}\Bbb{R}^4$ filled by all these pseudofibers is 
always dense --- meaning that the Legendre correspondence is ``almost surjective'' ---
and it can inverted on a dense subset. This contrasts strongly with situations
in classical mechanics (where the Legendre transform may degenerate, i.e.\,its image
may be reduced to a strict submanifold, in particular if the variational problem is
parametrization invariant) or in fields theory, if we restrict ourself to the de Donder--Weyl
theory or if we use the standard canonical approach (here the Legendre
transform degenerates as soon as we have a gauge invariance, a phenomenon known as
{\em Dirac's constraints}). A Hamiltonian function ${\cal H}$ can be defined on
${\cal M}:=\cup_{(x,y)\in \Bbb{R}^4}{\cal M}_{(x,y)}$ by setting
${\cal H}(x,y,P):= W(x,y,T,P)$, where $T$ is an implicit funtion of $(x,y,P)$
through the relation ${\partial W\over \partial T}(x,y,T,P) = 0$.\\

\noindent
The more convincing example is the trivial variational problem:
We just take $L =0$, so that any map from $\Bbb{R}^2$ to $\Bbb{R}^2$
is a critical point of our variational problem ! Then the image
${\cal M}_{(x,y)}$ of the Legendre correspondence is the union of the
complementary of the hyperplane $r=0$ and of $\{(e,p^\mu_i,r)=(e,0,0)/e\in \Bbb{R}\}$.
The Hamiltonian function is given by
${\cal H}(x,y,e,0,0) = e$ and 
\[
{\cal H}(x,y,e,p^\mu_i,r)=e - {p^1_1p^2_2 - p^1_2p^2_1\over r},
\]
if $r\neq 0$. 
One can then check that all Hamiltonian 2-curves are of the form
\[
\Gamma=\left\{\left(x,u(x),e(x)dx^1\wedge dx^2+ \epsilon_{\mu\nu}p^\mu_i(x)dy^i\wedge dx^\nu +
r(x) dy^1\wedge dy^2 \right)/x\in \Bbb{R}^2 \right\},
\]
where $u:\Bbb{R}^2\longrightarrow \Bbb{R}^2$ is an {\em arbitrary} smooth function,
$r:\Bbb{R}^2\longrightarrow \Bbb{R}^*$ is also an arbitrary smooth function and
\[
e(x) = r(x)\left( {\partial u^1\over \partial x^1}(x){\partial u^2\over \partial x^2}(x)
- {\partial u^1\over \partial x^2}(x){\partial u^2\over \partial x^1}(x)\right) + h,
\]
for some constant $h\in \Bbb{R}$, and
\[
p^\mu_i(x) = -r(x)\epsilon^{\mu\nu}\epsilon_{ij}
{\partial u^j\over \partial x^\nu}(x).
\]
Note that in this setting the de Donder--Weyl theory corresponds to the further
constraint $r=0$, which here implies that $p^\mu_i=0$: we thus recover the fact that
the Legendre transform completely degenerates.\\

\noindent
Other examples can be studied like for instance:\
\begin{itemize}
\item The harmonic map Lagrangian
$L(x,y,v) = {1\over 2}|v|^2$ where
$|v|^2:=(v^1_1)^2 + (v^1_2)^2 +(v^2_1)^2 +(v^2_2)^2$. Then one finds that
${\cal H}(q,p) = e + {1\over 1-r^2}
\left( {|p|^2\over 2} + r(p^1_1p^2_2 - p^1_2p^2_1)\right)$.\\
\item The Maxwell equations in two dimensions.
We take $L(x,y,v) = -{1\over 2}\left( v^1_2-v^2_1\right)^2$. Then
${\cal H}(q,p) = e + {(p^1_2 + p^2_1)^2 - 4p^1_1p^2_2\over 4r} 
-{1\over 4} {(p^1_2 - p^2_1)^2\over 2+r}$.
\end{itemize}
Beside the Legendre correspondence, the Lepage--Dedecker theory differs from the
de Donder--Weyl theory in many other aspects, sometime relatively subtle. For example,
on ${\cal M}:= \Lambda^nT^*{\cal N}$, with its standard multisymplectic form $\Omega$,
the set of algebraic observable $(n-1$)-forms coincides with the set of observable
$(n-1)$-forms. Hence $({\cal M}, \Omega)$ is an example of a pataplectic manifold.
I refer to our paper \cite{HeleinKouneiher2} for a complete exposition.\\

\noindent
Another question which is discussed in \cite{HeleinKouneiher1} and \cite{HeleinKouneiher2}
is the possibility and the relevance of considering observable forms of degree
$p-1$, where $p<n$. This was proposed first by I. Kanatchikov in \cite{Kanatchikov1},
\cite{Kanatchikov2}. For instance in the Hamiltonian system (\ref{2.hamilton/brut}),
one sees a disymmetry between $\pi^\mu_i$ and $u^i$ (which disappears when $n=1$):
the equations on $\pi^\mu_i$ involve a divergence whereas equations on $u^i$ prescribe
its derivatives in all direction. This reflects the fact that the $\pi^\mu_i$'s
are actually the components of an observable $(n-1)$-form, whereas $u^i$ could be
considered as an observable 0-form. Another beautiful example holds for gauge theories
(see \cite{Kanatchikov1}, \cite{HeleinKouneiher2}): where the gauge potential $A_\mu dx^\mu$
can be considered
as an observable 1-form, whereas the Faraday form $\star (dA+A\wedge A)$ as an
observable $(n-2)$-form. Moreover these forms are canonically conjugate (in a sense
similar to the duality between position and momentum variables in classical mechanics).
The first definition proposed in \cite{Kanatchikov1} (a $(p-1)$-form $F$
is observable if and only
if there exists an $(n-p)$-multivector $\xi$ such that $dF + \xi\iN \Omega = 0$) leads to a
quite interesting notion of graded Poisson bracket, but causes some difficulties when one tries
to generalize relation (\ref{2.dH}) to $(p-1)$-forms, for $1\leq p<n$. In
\cite{HeleinKouneiher2}, starting from another point of view, namely by characterising
the property which seems to be relevant in order to generalize relation (\ref{2.dH}),
we proposed the following alternative definition.
We define {\em collectively} the set of all observable $(p-1)$-forms,
for $1\leq p<n$: it is a vector subspace $\mathfrak{P}^*{\cal M}$ of the set of
sections of $\oplus_{p=1}^n\Lambda^{p-1}T^*{\cal M}$ such that
$\forall F_1,\cdots ,F_k\in \mathfrak{P}^*{\cal M}$, if
$dF_1\wedge \cdots \wedge dF_k$ is a section of $\Lambda^nT^*{\cal M}$, then
there exists a vector field $\xi$ on ${\cal M}$ such that
$dF_1\wedge \cdots \wedge dF_k + \xi\iN \Omega = 0$. This definition seems slightly
unpleasant at first glance (there could be several systems of observable forms,
i.e.\,several choices for $\mathfrak{P}^*{\cal M}$) but it provides the right hypothesis
in order to prove generalizations of (\ref{2.dH}) (see \cite{HeleinKouneiher2}).
Note also that it has some physical meaning (see the next paragraph).
One drawback is that it is now more delicate to define a notion of Poisson bracket
between such forms. However we were able to propose a partial definition which works for the most important cases (see \cite{HeleinKouneiher2}).\\

\noindent
Note that the above definition of observable $(p-1)$-forms collectively is an example of a mechanism by which observable quantities (and in particular space-time coordinates) could
merge out from intrinsic properties. In this spirit we also remarked in \cite{HeleinKouneiher2}
that the dynamical relation (\ref{2.dH}) can be replaced by a more general one:
\[
\{{\cal H},F\}dG_{|\Gamma} = \{{\cal H},G\}dF_{|\Gamma},
\]
(a generalisation for observable forms of lower degrees also exists). The 
underlying idea, that nothing can be measured in an absolute sense and that we can only compare
the results of two different measures, is very much in the spirit of general relativity. 

\section{Dynamical observable forms and perturbation theory}
\subsection{Dynamical observable $(n-1)$-forms and their motivations}
When building a quantum field theory, in order to achieve a relativistic invariance
and in particular to be free from any choice of time coordinate, one is led to use the
{\em Heisenberg point of view}. There a vector in the Hilbert (Fock) space of quantum states
represents the complete history over all space-time of a quantized field,
and observable operators act
on this Hilbert space with eigenvalues which are smeared integrals of functions of
the values of the fields
and their space-time derivatives on space-time. The classical counterpart of this point of
view, sometime called Einstein point of view, is to consider the set ${\cal E}$ of
solutions to the Euler--Lagrange equations of motion over all space-time to be the set
of physical states and to consider the set of functionals defined on ${\cal E}$ to be
the set of observables. If we wish to understand in a covariant way how to quantize these
fields it seems to be crucial to be able to define a Poisson bracket between observable functionals
and in particular between two observable functionals of the type $\int_{\Sigma}F:\Gamma\longmapsto
\int_{\Gamma\cap \Sigma}F$ and $\int_{\widetilde{\Sigma}}G:
\Gamma\longmapsto \int_{\Gamma\cap \widetilde{\Sigma}}G$,
even if $\Sigma$ and $\widetilde{\Sigma}$ are two different hypersurfaces. This is however
not clear in general. One possiblity arises when, for instance, $F$ is such that
\begin{equation}\label{4.dyn}
\{{\cal H},F\} = - d{\cal H}(\xi_F) = 0.
\end{equation}
We then say that $F$ is a {\em dynamical} observable $(n-1)$-form. Assume furthermore
that $\Sigma$ and $\widetilde{\Sigma}$ are homologous hypersurfaces, so that
$\widetilde{\Sigma} - \Sigma$ is the boundary of an open subset $D$. Then by applying
first Stokes' theorem and second (\ref{2.dH}) we obtain that
\[
\int_{\Gamma\cap \widetilde{\Sigma}}F - \int_{\Gamma\cap \Sigma}F = 
\int_{\Gamma\cap D}dF = \int_{\Gamma\cap D}\{{\cal H},F\}\omega = 0.
\]
Hence the two functionals $\int_{\Sigma}F$ and
$\int_{\widetilde{\Sigma}}F$ coincide on ${\cal E}$.
We can thus pose
\[
\left\{ \int_{\Sigma}F, \int_{\widetilde{\Sigma}}G\right\} :=
\left\{ \int_{\widetilde{\Sigma}}F, \int_{\widetilde{\Sigma}}G\right\}
= \int_{\widetilde{\Sigma}}\{F,G\}.
\]
Thus it remains to find dynamical observable forms. Here comes a surprise and a
relative deception. We quote H. Goldschmidt and S. Sternberg in \cite{GoldschmidtSternberg}:
``{\em For the free fields (i.e.\,quadratic Lagrangians) that arise in quantum field theory,
the algebra}\footnote{$P$ is here the quotient of the set of algebraic observable
$(n-1)$-forms by the subset of exact $(n-1)$-forms}
{\em $P$ is infinite dimensional and provides enough elements to yield the operators
of the associated free quantum fields. However computations done jointly with
S. Coleman to whom we are very grateful, seem to indicate that of $n\geq 3$, then for
``interacting Lagrangians'', i.e.\,those containing higher order terms, the algebra $P$
is finite dimensional, and hence does not provide enough operators for quantization}''.
A more detailed computation with basically the same conclusion
can be found in J. Kijowski's paper \cite{Kijowski1}. It is interesting here that this
question meet the same kind of difficulties as the quantization problem for fields:
the quantization procedure works when the classical equation is linear but fails
as soon as the problem become nonlinear ({\em interacting fields} in the language
of physicists). This is perhaps an indication that the two questions are related
(although there is no doubt that the quantization problem is {\em much more} difficult).\\

\noindent
There are however some ways to escape from this dead end. One is to remark that the set
of dynamical observable forms is roughly speaking in correspondence with the set of
symmetries of the problem (Noether theorem). In particular in the presence of a gauge symmetry we can produce
an infinite dimensional family of dynamical observable forms, which corresponds to the set of all current
densities smeared with any test function. We have discussed this approache in \cite{HeleinKouneiher2}.
It suggests the question whether a gauge symmetry could improve the quantization of a field
theory. In the same spirit it could be interesting to explore integrable systems, which possesses
infinitely many symmetries. A third possibility is when the problem is nonlinear but close to a linear
one: one can then hope to build observable functionals by perturbations. We expound here a simple
example which illustrates this idea.

\subsection{The interacting field: obstructions to dynamical observables forms}
We let $\eta_{\mu\nu}$ be a constant metric on $\Bbb{R}^n$ (which could be Euclidean or
Minkowskian), with inverse $\eta^{\mu\nu}$,
and we consider the following functional on the
set of maps $\varphi:\Bbb{R}^n\longrightarrow \Bbb{R}$:
\[
{\cal L}[\varphi]:= \int_{\Bbb{R}^n}\left( {1\over 2} \eta^{\mu\nu}
{\partial \varphi\over \partial x^\mu}
{\partial \varphi\over \partial x^\nu} + {m^2\over 2}\varphi^2 + {\lambda\over 3}\varphi^3\right)
\omega,
\]
where $\omega=dx^1\wedge \cdots \wedge dx^n$ and $\lambda$ is a scalar constant, that
we suppose to be small. Denoting by $\Delta:= - \eta^{\mu\nu}
{\partial ^2\over \partial x^\mu\partial x^\nu}$, the Euler--Lagrange equation is
\[
\Delta\varphi + m^2\varphi + \lambda \varphi^2 = 0.
\]
Since here the target space is 1-dimensional there is no difference between the
de Donder--Weyl theory and the other Lepage theories. The multisymplectic manifold ${\cal M}$
can hence either be constructed as the dual of the first affine jet bundle or with
$\Lambda^nT^*(\Bbb{R}^n\times \Bbb{R})$. We identify it with $\Bbb{R}^{2n+2}$, with the
coordinates $(x^\mu, \phi, e, p^\mu)$ and with the multisymplectic (or pataplectic) form
\[
\Omega:= de\wedge \omega + dp^\mu\wedge d\phi\wedge \omega_\mu,
\]
where $\omega_\mu:= {\vec{\partial \over \partial x^\mu}}\iN \omega$.
Then Hamiltonian $n$-curves are $n$-dimensional submanifolds $\Gamma$ of
${\cal M}\simeq \Bbb{R}^{2n+2}$ such that, for any point
$p\in \Gamma$ there exists a unique $n$-multivector $X$ tangent to $\Gamma$ at $p$ such that
$X\iN \Omega = (-1)^nd{\cal H}$, where
\[
{\cal H}(x,\phi,e,p):= e + {1\over 2}\eta_{\mu\nu}p^\mu p^\nu - {m^2\over 2}\phi^2
-{\lambda\over 3}\phi^3.
\]
The search for all dynamical algebraic observable $(n-1)$-forms consists in the following:
one looks at vector fields $\xi$ on ${\cal M}$ such that
\begin{equation}\label{4.pataplecto}
d\left( \xi\iN \Omega\right) = 0,
\end{equation}
which will implies that there exists an $(n-1)$-form $F$ such that $\xi=\xi_F$,
i.e.\,$dF+\xi\iN \Omega = 0$, and such that
\begin{equation}\label{4.dynamico}
\{{\cal H}, F\} = - d{\cal H}(\xi) = 0.
\end{equation}
This has been done in \cite{Kijowski1}. The results are that:
if $\lambda\neq 0$ the only solutions are $\xi = X^\mu{\vec{\partial \over \partial x^\mu}}$,
where $X^\mu$ are constants, if $\lambda=0$ the solutions are
$\xi = X^\mu{\vec{\partial \over \partial x^\mu}} + \eta^{\mu\nu}
{\partial \Phi\over \partial x^\nu}(x){\vec{\partial \over \partial p^\mu}}
+ \left(m^2\phi\Phi(x) - p^\mu{\partial \Phi\over \partial x^\mu}(x)\right)
{\vec{\partial \over \partial e}} + \Phi(x){\vec{\partial \over \partial \phi}}$, where
$\Phi$ is a solution of $\Delta\Phi+m^2\Phi=0$.\\

\noindent
Let us revisit partially this analysis: we assume for simplicity that 
$dx^\mu(\xi)=0$ (i.e.\,we throw away the $X^\mu$'s which correspond to parts of the
stress-energy-tensor). First one finds that such a $\xi$ satisfies (\ref{4.pataplecto})
if and only if
\begin{equation}\label{4.xi}
\xi= \left(P^\mu(x,\phi) - p^\mu
{\partial \Phi\over \partial \phi}(x,\phi)\right){\vec{\partial \over \partial p^\mu}}
+ \left(E(x,\phi) - p^\mu
{\partial \Phi\over \partial x^\mu}(x,\phi)\right){\vec{\partial \over \partial e}}
+ \Phi(x,\phi){\vec{\partial \over \partial \phi}},
\end{equation}
where $\Phi$, $E$ and $P^\mu$ are arbitrary functions of $(x,\phi)$ subject to the condition
\begin{equation}\label{4.compatible}
{\partial E\over \partial \phi} - {\partial P^\mu\over \partial x^\mu} = 0.
\end{equation}
Second the substitution of the value of $\xi$ in (\ref{4.dynamico}) leads to the system of equations
\begin{equation}\label{4.phi-et-P}
{\partial \Phi\over \partial \phi}(x,\phi)= 0\quad ,\quad \quad 
{\partial \Phi\over \partial x^\mu}(x,\phi) - \eta_{\mu\nu}P^\nu(x,\phi) =0,
\end{equation}
(which implies that $\Phi$ and $P^\mu$ depend only on $x$ and $P^\mu(x)=\eta^{\mu\nu}
{\partial \Phi\over \partial x^\nu}(x)$) and
\begin{equation}\label{4.impossible}
\left(m^2\phi + \lambda\phi^2\right) \Phi(x) - E(x,\phi) = 0.
\end{equation}
The system (\ref{4.phi-et-P}), (\ref{4.impossible}) has no nontrivial solution when
$\lambda\neq 0$, because by (\ref{4.compatible}) it would
contradict the fact that $P^\mu(x)=\eta^{\mu\nu}{\partial \Phi\over \partial x^\nu}(x)$.\\

\noindent
So let us forget condition (\ref{4.impossible}) and assume only (\ref{4.xi}),
(\ref{4.compatible}) and (\ref{4.phi-et-P}). By (\ref{4.compatible}) and (\ref{4.phi-et-P}) we deduce that
${\partial E\over \partial \phi}(x,\phi) = -\Delta\Phi(x)$ and so there exists a function
$A:\Bbb{R}^n\longrightarrow \Bbb{R}$ such that $E(x,\phi) = A(x)-\phi \Delta\Phi(x)$. We shall
assume $A=0$ in the following (since $A$ does not help in anything), we deduce that
\[
\xi = \eta^{\mu\nu}
{\partial \Phi\over \partial x^\nu}(x){\vec{\partial \over \partial p^\mu}}
- \left(\phi \Delta\Phi(x) + p^\mu{\partial \Phi\over \partial x^\mu}(x)\right)
{\vec{\partial \over \partial e}} + \Phi(x){\vec{\partial \over \partial \phi}},
\]
(then $\xi\iN \Omega = -dF$, where $F=\left( p^\mu\Phi(x) -\eta^{\mu\nu}\phi
{\partial \Phi\over \partial x^\nu}(x)\right)\omega_\mu$) and
\[
d{\cal H}(\xi) = - \phi\left( \Delta\Phi(x)+m^2\Phi(x)\right) - \lambda \phi^2\Phi(x).
\]
We now suppose that $\Phi=\Phi^{(1)}$, a solution of the equation
$\Delta\Phi^{(1)}+m^2\Phi^{(1)} = 0$. We denote by $\xi^{(1)}$ the corresponding
vector field and $F^{(1)}$ the associated observable $(n-1)$-form:
$F^{(1)} := \left( p^\mu\Phi^{(1)}(x) -\eta^{\mu\nu}\phi
{\partial \Phi^{(1)}\over \partial x^\nu}(x)\right)\omega_\mu$.
Then, instead of (\ref{4.dynamico}), we have
\begin{equation}\label{4.obstr0}
\{{\cal H},F^{(1)}\} = - d{\cal H}(\xi^{(1)}) = \lambda \phi^2\Phi^{(1)}(x),
\end{equation}
and thus
\[
\int_{\Gamma\cap \partial D}F^{(1)} = \int_{\Gamma\cap D}dF^{(1)} =
\int_{\Gamma\cap D}\lambda \phi^2\Phi^{(1)}(x)\omega.
\]

\subsection{Second order correction}
We now add to the functional $\int_{\partial D} F^{(1)}$ another functional of the form
\[
\begin{array}{cccc}
\displaystyle \lambda
\int_{\partial D}\int_{\partial D}F^{(2)}: & {\cal E} & \longrightarrow & \Bbb{R}\\
 & \Gamma & \longmapsto & \displaystyle 
\lambda\int_{\Gamma\cap \partial D}\int_{\Gamma\cap \partial D}F^{(2)},
\end{array}
\]
where $F^{(2)}\in \Gamma({\cal M},\Lambda^{n-1}T^*{\cal M}) \otimes
\Gamma({\cal M},\Lambda^{n-1}T^*{\cal M})$ (here $\Gamma({\cal M},\Lambda^{n-1}T^*{\cal M})$
is the set of sections over ${\cal M}$ of the bundle $\Lambda^{n-1}T^*{\cal M}$, i.e.\,the set
of $(n-1)$-forms over ${\cal M}$). We shall make the following hypotheses on $F^{(2)}$:
we let $\Omega^{\otimes 2}:= \Omega\otimes \Omega$ be in
$\Gamma({\cal M},\Lambda^{n+1}T^*{\cal M}) \otimes \Gamma({\cal M},\Lambda^{n+1}T^*{\cal M})$
and we assume that there exists a ``bivector'' $\xi^{(2)}$, i.e.\,an element of
$\Gamma({\cal M},T{\cal M}) \otimes \Gamma({\cal M},T{\cal M})$, such that
\begin{equation}\label{4.dF2}
d^{\otimes 2}F^{(2)} = (-1)^2\xi^{(2)}\iN \!\!\!\!\!^2\;\; \Omega\otimes \Omega.
\end{equation}
It deserves some definitions about notations. Let us abbreviate by $z^I$ the system of coordinates
$(x^\mu,y^i,e,p^\mu)$ on ${\cal M}$ and by $(z^I_1;z^J_2)=(x_1^\mu,y_1^i,e_1,p_1^\mu;x_2^\mu,y_2^i,e_2,p_2^\mu)$
coordinates on ${\cal M}\times {\cal M}$.
Then any $d^{\otimes 2}$ is the unique linear operator from
$\Gamma({\cal M},\Lambda^{n-1}T^*{\cal M}) \otimes
\Gamma({\cal M},\Lambda^{n-1}T^*{\cal M})$ to
$\Gamma({\cal M},\Lambda^nT^*{\cal M}) \otimes
\Gamma({\cal M},\Lambda^nT^*{\cal M})$ such that if
$\alpha^{(2)}\in \Gamma({\cal M},\Lambda^{n-1}T^*{\cal M}) \otimes
\Gamma({\cal M},\Lambda^{n-1}T^*{\cal M})$ writes
\[
\alpha^{(2)} = 
\alpha(z_1^I,z_2^J)
dz_1^{I_1}\wedge \cdots \wedge dz_1^{I_{n-1}}\otimes
dz_2^{J_1}\wedge \cdots \wedge dz_2^{J_{n-1}},
\]
then 
\[
d^{\otimes 2}\alpha^{(2)}:= \sum_{I,J} 
{\partial ^2\alpha \over \partial z_1^I \partial z_2^J}(z_1^I,z_2^J)
dz_1^I\wedge dz_1^{I_1}\wedge \cdots \wedge dz_1^{I_{n-1}}\otimes
dz_2^J\wedge dz_2^{J_1}\wedge \cdots \wedge dz_2^{J_{n-1}}.
\]
(Remark: similar tensor product of differential forms and operators $d^{\otimes 2}$
were used in \cite{Iso} for the purpose of proving isoperimetric inequalities through
calibrations.)
Similarly $\iN \!\!\!\!\!^2\;\;$ is an operation with standard linear properties such that
if $\xi^{(2)}\in \Gamma({\cal M},T{\cal M}) \otimes \Gamma({\cal M},T{\cal M})$ writes
\[
\xi^{(2)} = 
\xi(z_1^I,z_2^J)
{\vec{\partial \over \partial z_1^I}} \otimes {\vec{\partial \over \partial z_2^J}},
\]
then
\[
\xi^{(2)}\iN \!\!\!\!\!^2\;\; \Omega^{\otimes 2}:= 
\xi(z_1^I,z_2^J)\left({\vec{\partial \over \partial z_1^I}}\iN \Omega\right)
\otimes \left({\vec{\partial \over \partial z_2^J}}\iN \Omega\right).
\]
Then, denoting by $d_1:=d\otimes Id$ and $d_2:=Id\otimes d$, so that $d^{\otimes 2}
=d_1\circ d_2$, we have
\[
\begin{array}{ccl}
\displaystyle 
\int_{\Gamma\cap \partial D}\int_{\Gamma\cap \partial D}F^{(2)} & = &\displaystyle 
\int_{\Gamma\cap D}\int_{\Gamma\cap \partial D}d_1F^{(2)} = 
\int_{\Gamma\cap D}\int_{\Gamma\cap D}d^{\otimes 2}F^{(2)}\\
 & = & \displaystyle 
\int_{\Gamma\cap D}\int_{\Gamma\cap D}\xi^{(2)}\iN \!\!\!\!\!^2\;\;
\Omega\otimes \Omega \\
 & = & \displaystyle 
\int_{\Gamma\cap D}\int_{\Gamma\cap D}\left(\xi^{(2)}\iN \!\!\!\!\!^2\;\;
\Omega\otimes \Omega\right)\left( X(z_1)\otimes X(z_2)\right)\omega\otimes \omega \\
\\
 & & (X(z)\hbox{ is there a }n\hbox{-multivector tangent to }\Gamma\hbox{ at }z
\hbox{, s.t.\,}\omega_z(X(z)) = 1)\\
\\
 & = & \displaystyle 
\int_{\Gamma\cap D}\int_{\Gamma\cap D}(-1)^{2n}\left(X(z_1)\otimes X(z_2)\iN \!\!\!\!\!^2\;\;
\Omega\otimes \Omega\right)\left( \xi^{(2)}\right)\omega\otimes \omega \\
 & = & \displaystyle 
\int_{\Gamma\cap D}\int_{\Gamma\cap D}d{\cal H}_{z_1}\otimes d{\cal H}_{z_2}
\left( \xi^{(2)}\right)\omega\otimes \omega,
\end{array}
\]
where  we have used the fact that
$\left(X(z_1)\otimes X(z_2)\iN \!\!\!\!\!^2\;\;
\Omega\otimes \Omega\right)_{|\Gamma\times \Gamma} = (-1)^{2n}
\left(d{\cal H}_{z_1}\otimes d{\cal H}_{z_2}\right)_{|\Gamma\times \Gamma}$.
The idea is to look for an $F^{(2)}$ such that
\begin{equation}\label{4.goal2}
\{{\cal H}^{\otimes 2},F^{(2)}\}:=
d{\cal H}_{z_1}\otimes d{\cal H}_{z_2}\left( \xi^{(2)}\right) = 
- \Phi^{(1)}(x_1)\delta(x_1-x_2)\phi_1\phi_2 + {\cal O}(\lambda),
\end{equation}
where $\delta$ is the Dirac distribution on $\Bbb{R}^n$. Then
\[
\lambda\int_{\Gamma\cap \partial D}\int_{\Gamma\cap \partial D}F^{(2)}
= - \lambda\int_{\Gamma\cap D}\Phi^{(1)}(x)\phi^2\omega + {\cal O}(\lambda^2),
\]
so that $\int_{\Gamma\cap \partial D}F^{(1)} + \lambda\int_{\Gamma\cap \partial D}\int_{\Gamma\cap \partial D}F^{(2)} = {\cal O}(\lambda^2).$\\

\noindent
We choose $F^{(2)}$ of the form
\[
F^{(2)}=\left( p_1^\mu - \eta^{\mu\lambda}\phi_1{\partial \over \partial x_1^\lambda}\right)
\left( p_2^\nu - \eta^{\nu\sigma}\phi_1{\partial \over \partial x_2^\sigma}\right)
\Phi^{(2)}(x_1,x_2)\omega_\mu\otimes \omega_\nu,
\]
where $\Phi^{(2)}:\Bbb{R}^n\times \Bbb{R}^n\longrightarrow \Bbb{R}$ is a function
to be precised later. One can check that such an $F^{(2)}$ satisfies (\ref{4.dF2}) with\\

\noindent
$\displaystyle \xi^{(2)} = 
\left( \left( p_1^\mu{\partial \over \partial x_1^\mu} + \phi_1\Delta_1\right)
	{\vec{\partial \over \partial e_1}}
	- \eta^{\mu\lambda}{\partial \over \partial x_1^\lambda} 
	{\vec{\partial \over \partial p_1^\mu}}
	- {\vec{\partial \over \partial \phi_1}} \right)$
\[
	\otimes
\left( \left( p_2^\nu{\partial \over \partial x_1^\nu} + \phi_2\Delta_2\right)
	{\vec{\partial \over \partial e_2}}
	- \eta^{\nu\sigma}{\partial \over \partial x_2^\sigma}
	{\vec{\partial \over \partial p_2^\nu}}
	- {\vec{\partial \over \partial \phi_2}} \right)
\Phi^{(2)}(x_1,x_2).
\]
Here we denote by $\Delta_1:= \eta^{\mu\nu}{\partial ^2\over \partial x_1^\mu\partial x_1^\nu}$
and $\Delta_2:= \eta^{\mu\nu}{\partial ^2\over \partial x_2^\mu\partial x_2^\nu}$ and we have
introduced a symbolic notation in order to shorten the expression of $\xi^{(2)}$
(which is quite long to write): one should understand that this expression should be
developp using the rule $\left(K_1\vec{\partial \over \partial z_1^I}\otimes
K_2\vec{\partial \over \partial z_2^J}\right)A(x_1,x_2) = 
\left(K_1K_2A(x_1,x_2)\right)\vec{\partial \over \partial z_1^I}\otimes
\vec{\partial \over \partial z_2^J}$, for all linear differential operator
$K_1$ (resp. $K_2$) acting on the variables $x_1^\mu$ (resp. $x_2^\mu$)
(for instance $K_1$ can be $p_1^\mu{\partial \over \partial x_1^\mu} + \phi_1\Delta_1$,
${\partial \over \partial x_1^\lambda}$ or $1$).\\

\noindent
Then one compute that
\[
\begin{array}{ccl}
\{{\cal H}^{\otimes 2},F^{(2)}\} =
d{\cal H}_{z_1}\otimes d{\cal H}_{z_2}\left( \xi^{(2)}\right) & = &
\;\phi_1\phi_2\left(\Delta_1+m^2\right)\left(\Delta_2+m^2\right) \Phi^{(2)}(x_1,x_2)\\
 & & + \,\lambda\left( \phi_1^2\phi_2\left(\Delta_2+m^2\right)
+ \phi_1\phi_2^2\left(\Delta_1+m^2\right)\right)\Phi^{(2)}(x_1,x_2)\\
 & & +\, \lambda^2\phi_1^2\phi_2^2\Phi^{(2)}(x_1,x_2).
\end{array}
\]
Thus in order to achieve (\ref{4.goal2}) it suffices to choose $\Phi^{(2)}$ such that
\[
\left(\Delta_1+m^2\right)\left(\Delta_2+m^2\right) \Phi^{(2)}(x_1,x_2) = 
- \Phi^{(1)}(x_1)\delta(x_1-x_2).
\]
A solution to this equation is {\em formally}
\[
\Phi^{(2)}(x_1,x_2) = - \int_{\Bbb{R}^n}\Phi^{(1)}(t)G(t,x_1)G(t,x_2)dt,
\]
where $G$ is the Green function on $\Bbb{R}^n$ of the operator $\Delta+m^2$.

\subsection{Perturbation series}
We have produced two observables, $\int_{\partial D}F^{(1)}$ and
$\int_{\partial D}F^{(1)} + \lambda\left(\int_{\partial D}\right)^2F^{(2)}$,
which are approximately vanishing, up to order $\lambda$ (resp. $\lambda^2$).
This looks clearly as the beginning of an infinite expansion series
defining the functional
\begin{equation}\label{5.infini}
\Gamma\longmapsto 
\sum_{k=1}^{\infty} \lambda^{k-1}\left( \int_{\Gamma\cap \partial D}\right)^k F^{(k)}.
\end{equation}
Here we let $F^{(k)}:= \left[\prod_{j=1}^k\left( 
p^\mu_j-\eta^{\mu\lambda}\phi_j{\partial \over \partial x_j^\lambda}\right)\right]
\Phi^{(k)}(x_1,\cdots ,x_k)\in \Gamma({\cal M},\Lambda^{n-1}T^*{\cal M})^{\otimes k}$,
where $\Phi^{(k)}$ is a function on $\left( \Bbb{R}^n\right)^k$. We need to choose
functions $\Phi^{(k)}$ in such a way that
\[
\sum_{k=1}^\infty \lambda^{k-1}\left(\int_{\Gamma\cap \partial D}\right)^k
\{{\cal H}^{\otimes k},F^{(k)}\} \omega^{\otimes k}= 0,
\]
where $\{{\cal H}^{\otimes k},F^{(k)}\}:= (-1)^kd^{\otimes k}{\cal H}^{\otimes k}(\xi^{(k)})
= \left[\prod_{j=1}^k\left( \phi_j(\Delta_j+m^2)+\lambda\phi_j^2\right)\right]
\Phi^{(k)}(x_1,\cdots ,x_k)$.
Of course the computation of the further terms should more and more complicated but
is possible in principe and should be
described by graphs analog to Feynman's diagramm. Note that these graphs should all
be trees (i.e.\,without loops), since they describe classical observable functionals.\\

\noindent
Then (\ref{5.infini}) provides us with a vanishing functional, if $\lambda$ is sufficiently
small. This was not exactly our original motivation, which was to find non vanishing
dynamical functionals. These can be obtained as follows. Assume for instance that
$\eta^{\mu\nu}$ is Minkowskian, i.e.\,$\Delta$ is hyperbolic, and $\partial D = \widetilde{\Sigma}
- \Sigma$, where $\Sigma$ is a fixed space-like hypersurface (say $\{x^0=t_0\}$ for
some fixed $t_0$) and $\widetilde{\Sigma}$ is a parallel space-like hypersurface
(say $\{x^0=t\}$, where $t\neq t_0$).
Then we prescribe all functions $\Phi^{(k)}$, for $k\geq 2$, in
such a way that they vanish and their first time derivative vanish along $\Sigma$.
This implies that
\[
\sum_{k=1}^{\infty} \lambda^{k-1}\left( \int_{\Gamma\cap \partial D}\right)^k F^{(k)}
= \sum_{k=1}^{\infty} \lambda^{k-1}\left( \int_{\Gamma\cap \widetilde{\Sigma}}\right)^k F^{(k)}
- \int_{\Gamma\cap \Sigma} F^{(1)}.
\]
Hence we get the coincidence of the two functionals
$\sum_{k=1}^{\infty} \lambda^{k-1}\left( \int_{\widetilde{\Sigma}}\right)^k F^{(k)}$
and $\int_\Sigma F^{(1)}$ on ${\cal E}$. Then it remains of course to define the
bracket $\left\{ \sum_{k=1}^{\infty} \lambda^{k-1}\left( \int_{\widetilde{\Sigma}}\right)^k F^{(k)},
\int_{\widetilde{\Sigma}} G\right\}$, in order to set
\[
\left\{ \int_\Sigma F^{(1)},
\int_{\widetilde{\Sigma}} G\right\}:=
\left\{ \sum_{k=1}^{\infty} \lambda^{k-1}\left( \int_{\widetilde{\Sigma}}\right)^k F^{(k)},
\int_{\widetilde{\Sigma}} G\right\}.
\]
In principle there should be no difficulty in defining the above Poisson bracket, either
by coming back to the Poisson bracket obtained by the standard canonical theory of
physicists or by using for instance the theory expounded in \cite{Deligne-Freed}.
Note that in order to make connection with quantum field theory, which provides us
quantum scattering amplitudes, it is more appropriate to choose the slice
$\Sigma$ at infinity, i.e.\,such that $t_0=-\infty$ or $t_0=\infty$.\\

\noindent
Lastly we also remark that replacing the set of functionals $\int_\Sigma F$ by the set
of functionals of the type (\ref{5.infini}) has another advantage: it is then possible
to define a product law between such functionals, by the rule
\[
\left(\sum_{k=1}^{\infty} \left( \int_{\Sigma}\right)^k F^{(k)}\right)
\left(\sum_{k=1}^{\infty} \left( \int_{\Sigma}\right)^k G^{(k)}\right)
= \sum_{k=1}^{\infty}\sum_{l=0}^k\left( \int_{\Sigma}\right)^k
F^{(l)}\otimes G^{(k-l)}.
\]

\section{Conclusion}
We have tried here to introduce the Reader to multisymplectic formalisms, mainly
through the de Donder--Weyl theory, we have explained quickly a more general
framework based on Lepage--Dedecker developped in \cite{HeleinKouneiher1} and
\cite{HeleinKouneiher2} (that is, we believe, supported by a more relativistic
point of view) and we have explained how a perturbative theory analog to the
theory of Feynman and Schwinger could be be built for classical solutions. This
theory need of course to be developped and it should be interesting to understand
whether it could lead to the perturbative quantum field theory by a direct quantization.\\

\noindent
We have not discussed many other important questions like the construction
of a non perturbative quantum field theory in this framework (interesting results
have been obtained by I. Kanatchikov, see \cite{Kanatchikov3}), or how such theories could
help in understanding hyperbolicity of variational nonlinear hyperbolic partial 
differential equations as done in D. Christodoulou's book \cite{Christodoulou}.
We should also mention the existence of other covariant theories as for example
F. Takens' one \cite{Takens} (see also \cite{Zuckerman} or \cite{Deligne-Freed}).\\

\noindent
{\em Acknowledgements --- I thank Joseph Kouneiher for useful discussions and comments
on this text.}

\noindent
{\it Fr\'ed\'eric H\'elein,\\
CMLA, ENS de Cachan} and {\it Institut Universitaire de France\\
64, avenue du Pr\'esident Wilson,\\
94235 Cachan Cedex, France\\
helein@cmla.ens-cachan.fr\\
http://www.cmla.ens-cachan.fr/Utilisateurs/helein/}


\begin{thebibliography}{99}

\newcommand{\bib}[1]{\bibitem{#1}}

\bib{Binz} E. Binz, J. \'{S}niatycki and  H. Fisher, 
{\sl Geometry of Classical Fields, } (North-Holland, Amsterdam, 1989) 

\bib{Boerner} H. Boerner, {\em Carath\'eodory's Eingang zur Variationsrechnung},
Jber. dt. Math.-Vereinig. 56 (1953), 31--58; 
{\em Darstellung von Gruppen}, Springer, Berlin--G\"ottingen--Heidelberg
(1955). 

\bib{Born} M. Born, {\em On the quantum theory of the electromagnetic field},
Proc. Roy. Soc. London A143 (1934), 410--437.


\bib{Cara} C. Carath\'eodory, 
{\em \"Uber die Extremalen und geod\"atischen Felder in der 
Variationsrechnung der mehrfachen Integrale, } 
Acta Sci. Math. (Szeged) 4 (1929) 193-216.

\bib{Cartan} E. Cartan, {\em Les espaces m\'etriques fond\'es sur la notion
d'aire}, 1933.

\bib{Christodoulou} D. Christodoulou, {\em The action principle and partial differential
equations}, Annals of Mathematics Studies, 146, Princeton University Press 2000.

\bib{Dedecker} P. Dedecker, {\em Calcul des variations, formes diff\'erentielles et
champs g\'eod\'esiques}, in {\em G\'eom\'etrie diff\'erentielle}, Colloq. Intern. du CNRS LII,
Strasbourg 1953, Publ. du CNRS, Paris, 1953, p. 17-34; {\em On the generalization of
symplectic geometry to multiple integrals in the calculus of variations}, in {\em Differential
Geometrical Methods in Mathematical Physics}, eds. K. Bleuler and A. Reetz, Lect. Notes Maths.
vol. 570, Springer-Verlag, Berlin, 1977, p. 395-456.

\bib{dDW} T. de Donder, {\em Th\'eorie invariante du calcul des variations}, Nuov.
\'ed. (Gauthiers--Villars, Paris 1935).

\bib{Deligne-Freed} P. Deligne, D. Freed,
{\em Classical field theory}, in {\it Quantum fields and strings: a
course for mathematicians, Volume 1}, P. Deligne, P. Etingof, D.S. Freed, L.C. Jeffrey,
D. Kazhdan, J.W. Morgan, D.R. Morrison and E. Witten, editors, American Mathematical
Society, 1999.

\bib{Echeverria}
A. Echeverria-Enriquez, M. Mu\~ noz-Lecanda, N. Roman-Roy, {\em Multivector
field formulation of Hamiltonian field theories}, J.Phys.A, v.32 (1999)
8461, math-ph/9907007

\bib{GarciaPerez} P.L. Garc\'\i a, A. P\'erez-Rend\'on, {\em Symplectic approach to the
theory of quantized fields, II}, Archive Rat. Mech. Anal. 43 (1971), 101--124.

\bib{Gawedski} K. Gaw\c{e}dski, 
{\em On the generalization of the canonical formalism in the 
classical field theory, }
Rep. Math. Phys. 3 (1972) 307-326.


\bib{GiaquintaHildebrandt} M. Giaquinta, S.Hildebrandt, {\em Calculus of variations}, Vol.
1 and 2, Springer, Berlin 1995 and 1996.

\bib{GoldschmidtSternberg} H. Goldschmidt, S. Sternberg, {\em The Hamilton--Cartan
formalism in the calculus of variations}, Ann. Inst. Fourier 23, p. 203--267 (1973).

\bib{Gotay1} M.J. Gotay,
{\em An exterior differential systems approach to the Cartan form, }  
 in {\sl Symplectic Geometry and Mathematical Physics, } 
eds. P. Donato, C. Duval, e.a. (Birkh\"{a}user, 
Boston, 1991) p. 160-188


\bib{Gotay2} M.J. Gotay, 
{\em A multisymplectic framework for classical field theory 
and the calculus of variations I. Covariant Hamiltonian formalism, }
in {\sl Mechanics. Analysis and Geometry: 
200 Years after Lagrange}, ed. M. Francaviglia (North Holland, 
Amsterdam, 1991) p. 203-235

\bib{Gotay3} M.J. Gotay, 
{\em A multisymplectic framework for classical field theory 
and the calculus of variations II. Space + time decomposition, } 
Diff. Geom. and its Appl. 1 (1991) 375-390

\bib{gimmsy} M.J. Gotay, J. Isenberg, J. Marsden (with the collaboration of
R. Montgomery, J.\'Sniatycki, P.B. Yasskin),
{\em Momentum  Maps and Classical Relativistic Fields. Part I: covariant field
theory}, arXiv:physics/9801019.

\bib{Iso}F. H\'elein, {\em In\'egalit\'e isop\'erim\'etrique et calibration}, Ann.
Inst. Fourier 44, fasc. 4 (1994), 1211--1218; {\em Isoperimetric inequalities and calibrations},
in {\em Progress in partial differential equations: the Metz surveys 4}, M. Chipot and
I. Shafrif ed., Pitman Research Notes in Mathematics Series 345, Longman 1996.

\bib{HeleinKouneiher1} F. H\'elein, J. Kouneiher, {\em Finite dimensional
Hamiltonian formalism for gauge and quantum field theory}, J. Math. Physics,
vol. 43, No. 5 (2002).

\bib{HeleinKouneiher2} F. H\'elein, J. Kouneiher, {\em Covariant Hamiltonian formalism
for the calculus of variations with several variables}, preprint 2002, CMLA, ENS de Cachan,
arXiv:math-ph/0211046.



\bib{Kanatchikov1}
I. V. Kanatchikov {\em Canonical structure of classical field theory in the polymomentum phase space},
hep-th/9709229

\bib{Kanatchikov2}
I. V. Kanatchikov, {\em On the canonical structure of the De Donder-Weyl covariant Hamiltonian formulation of field theory I.
Graded Poisson brakets and the equation of motion}, hep-th/9312162

\bib{Kanatchikov3}
I. V. Kanatchikov, {\em Covariant geometric prequantization of fields}, 
(arXiv:gr-qc/0012038)Proc. 19th Marcel Grossmann Meeting, Rome (Italy), July 2000,
World Scientific, Singapore 2000; {\em Precanonical quantization and the Schr\"odinger wave
functional}, Phys. Letters A 283 (2001), 25--36.

\bib{Kastrup} H. Kastrup, 
{\em Canonical theories of Lagrangian dynamical systems in physics, }
Phys. Rep. 101 (1983) 1-167 

\bib{Kijowski1} J. Kijowski, {\em A finite dimensional canonical formalism in the classical
field theory}, Comm. Math. Phys. 30 (1973), 99-128.

\bib{Kijowski2} J. Kijowski, {\em Multiphase spaces and gauge in the
calculus of variations}, Bull. de l'Acad. Polon. des Sci., S\'erie sci. Math., Astr. et
Phys. XXII (1974), 1219-1225.

\bib{KijowskiSzczyrba} J. Kijowski, W. Szczyrba, {\em A canonical structure for classical
field theories}, Comm. Math. Phys. 46(1976), 183-206.

\bib{KijowskiTulczyjew} J. Kijowski, W.M. Tulczyjew, {\em A symplectic
framework for field theories}, Springer-Verlag, Berlin, 1979.

\bib{Lepage} T.H.J. Lepage, {\em Sur les champs g\'eod\'esiques du calcul
des variations}, Bull. Acad. Roy. Belg. Cl. Sci. V. S\'er. 22 (1936), 716--729; 1036--1046.

\bib{Martin} D. H. Martin,
{\em Canonical variables and geodesic fields for the calculus of
variations of multiple integrals in parametric form}, Math. Z. 104 (1968), 16-27.


\bib{Rund} H. Rund, {\em The Hamilton-Jacobi Theory in the Calculus of 
Variations}, D. van Nostrand Co. Ltd., Toronto, etc. 1966
(Revised and augmented reprint, Krieger Publ., New York, 1973).


\bib{Sardanashvily}
G. Sardanashvily, {\em Generalized Hamiltonian Formalism for Field Theory}, ed.
World Scientific, Singapore, 1995.

\bib{Snyatycki} J. \'Snyatycki, {\em On the geometric structure of classical field theory
in Lagrangian formulation}, Proc. Camb. Phil. Soc. 68 (1970), 475--483.

\bib{Takens} F. Takens, {\em A global version of the inverse problem of the calculus
of variations}, J. Diff. Geom. 14 (1979), 543--569.

\bib{Tulczyjew1} W.M. Tulczyjew, Warsaw seminar in {\em Geometry of phase space}, 
unpublished, 1968.


\bib{Weyl} H. Weyl, 
{\em Geodesic fields in the calculus of variations, } 
Ann. Math. (2)  36 (1935) 607-629.

\bib{Zuckerman} G.J. Zuckerman, {\em Action principles and global geometry}, Mathematical
aspects of string theory, ed. S.T. Yau, World Scientific Publishing, 1987, 259--284.

\end{thebibliography}
\end{document}